\def\tc{$T_c$\ }     
\def\ba{$\rm Ba(Fe_{1-x}Co_x)_2As_2$\ }  %
\def\dtc{$\delta T_c$}
\def\dl{$\delta l$}
\def\ha{$\mu_0H_a$\ }     
\def\hir{$H_{irr}$\ }     
\def\jc{$j_c$\ }     

\documentclass[prb,twocolumn]{revtex4}
\usepackage{graphicx}
\usepackage{amssymb,amsmath}

\begin{document}

\title{Microwave absorption study of pinning regimes in $\bf Ba(Fe_{1-x}Co_x)_2As_2$ single crystals}

\author{Yu. Talanov$^1$}\email{talanov@kfti.knc.ru} 
\author{N. Beisengulov$^1$} 
\author{G. Kornilov$^1$}
\author{T. Shaposhnikova$^1$}
\author{E. Vavilova$^1$}
 \affiliation{$^{1}$Zavoisky Physical-Technical Institute, 420029, Kazan, Russia}
 
\author{C. Nacke$^2$} 
\author{S. Wurmehl$^2$} 
\author{N. Panarina$^2$} 
\author{C. Hess$^2$} 
\author{V. Kataev$^2$} 
\author{B. B\"{u}chner$^{2,3}$}
 \affiliation{$^{2}$Leibniz Institute for Solid State and Materials Research IFW Dresden, D-01171 Dresden, Germany}
 \affiliation{$^{3}$Institut f\"{u}r Festk\"{o}rperphysik, Technische Universit\"{a}t Dresden, D-01062 Dresden, Germany}

\date{\today}

\begin{abstract}
Magnetic field dependent modulated microwave absorption (MMWA) measurements have been carried out to investigate vortex pinning effects in single crystals of the iron-based high-\tc superconductor \ba  with three different cobalt doping levels of $x = 0.07$, 0.09, and 0.11. The dependence of the MMWA hysteresis loops on temperature, magnetic field, and Co-concentration have been measured and analyzed using a theoretical model of microwave absorption in superconductors. The analysis reveals that in an underdoped crystal ($x = 0.07$) the so called \dtc-pinning due to magnetically ordered regions defines the temperature dependence of the critical current density, while in the optimally doped ($x = 0.09$) and overdoped ($x = 0.11$) samples the pinning is governed by structural imperfections due to inhomogeneous distribution of cobalt dopant and has the so called \dl\ character. 
\end{abstract}

 \maketitle

\section{Introduction}
The \ba compound is likely the most investigated material among all known superconducting pnictides. Such an interest can be explained by both unusual physical properties inherent in this compound and availability of high quality single crystals. For the crystal growth most commonly a self-flux or the Bridgman techniques are used \cite{Aswartham11}. Crystals grown by different techniques may vary in properties, in particular, in the pinning strength and, as a result, in the superconducting critical current density. The vortex pinning effects in the \ba single crystals have been studied by transport, magnetic, magneto-optic and relaxation measurements (see, for example, Refs. \cite{Prozorov08, Prozorov09, Yamamoto09,  Prozorov09b}). The dependence of the pinning strength on environmental conditions (temperature $T$ and magnetic field \ha) and the cobalt doping level has been investigated. It has been established that the \ba samples have a large critical current density $j_c$ \cite{Prozorov08,Yamamoto09} and a steep slope of the irreversibility line $H_{irr}(T)$\ \cite{Yamamoto09}. A symmetric shape of the pinning strength function $F_p$\ versus field $H$\ allowed the authors of Ref. \cite{Yamamoto09} to suggest the presence of a strong pinning nanostructure. By simultaneous analysis of the magnetization curves $M(H)$ and the domain structure, the authors of Ref. \cite{Prozorov09b} concluded that the twin boundaries play a role of such nanostructure.

In the present work we have investigated \ba crystals with different Co concentration (from underdoped to overdoped) using the magnetic field dependent modulated microwave absorption (MMWA) measurements. We have studied the evolution of the MMWA signal as a function of temperature, applied magnetic field and doping level. The relevant superconducting parameters, in particular, the critical current density have been estimated through numerical simulation of the experimental MMWA loops. Furthermore, the MMWA method used in the present work has enabled, beyond the scope of the static magnetization measurements, not only to estimate the critical current density, but also to obtain additional insights into other effects, in particular on the influence of the thermal fluctuations and on the pinning center density. The theoretical analysis enables a distinction of pinning regimes in the studied single crystals with different Co-doping, as well as to establish the origin of pinning centers. In particular, we have found clear indications for pinning of vortices on magnetically ordered nanoscale regions in the underdoped sample, whereas at higher Co doping the pinning is related to local structural disorder and imperfections.    

\section{Experimental}

\ba single crystals with the Co content $x = 0.07$, 0.09 and 0.11 were grown by the vertical Bridgman technique. The preparation procedure is described in detail in Ref. \cite{Aswartham11}.  The Co concentration was determined by elemental analysis using energy dispersive X-ray (EDX) mode of a scanning electron microscope (see Ref.\cite{Aswartham11} for details). All $x$\ values referred to in this work are the Co contents as revealed by EDX.  Among all crystals the one with $x = 0.09$ has the maximal superconducting critical temperature $T_c = 26$~K (Fig. 1 and Table 1) and thus is considered as an optimally doped (OP) sample. Respectively, the crystal with $x = 0.07$\ is denoted as underdoped (UD), and the crystal with $x = 0.11$\ as overdoped (OD). In literature the optimal cobalt concentration $x$ for superconductivity in this series is reported to lie in the range from 6\% to 10\% depending on the method and regimes of preparation \cite{Aswartham11,Lester09,Chu09,Ning09,Wang09}.  Note that for Co concentration less than optimal two areas in the phase diagram $T$\ vs $x$\ can be distinguished: the one corresponding to the spin density wave (SDW) state (at high temperature) and the other corresponding to the superconducting state (at low temperatures) \cite{Aswartham11}. According to a neutron scattering study \cite{Lester09}, at these concentrations at $T<T_c$ the superconducting phase coexists with the magnetically ordered phase.

The parameters of the superconducting transition were determined from the temperature dependence of AC susceptibility (Fig. 1). The obtained values are summarized in Table 1. The onset temperature $T_c^{on}$\ was determined as a temperature at which the AC susceptibility signal starts to deviate from the zero line. $T_c^m$\ is the temperature at the half-height of the transition. The width of the transition  $T_c$\ is determined as the interval between the points of 10\% and 90\% of the total transition height. Notably the UD sample has the broadest transition. 

\begin{figure}
\includegraphics[width=7.5cm]{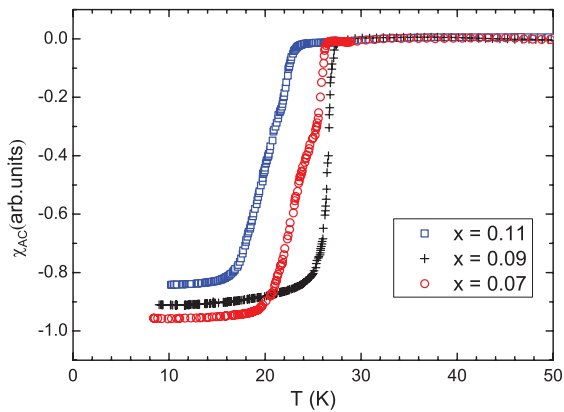}
\caption{\label{trans} Temperature dependence of AC-susceptibility for three $\rm Ba(Fe_{1-x}Co_x)_2As_2$\ samples with different Co concentration. }
\end{figure}

 There is no indication of the magnetic phase transition in the temperature dependence of the AC susceptibility in the studied samples. However for the UD single crystal it is observed in the resistivity $\rho(T)$\ measurements (Fig. 2) which turn out to be a very sensitive probe of the magnetic phase transition manifesting in a clear minimum of the derivative curve $d[\rho(T)]/dT$(for details see \cite{Aswartham11}). In fact, following the procedure discussed in Ref.\cite{Aswartham11}, besides a transition to the magnetically ordered phase of SDW type at $T_{SDW} = (40 \pm 2)$~K corresponding to a minimum of $d[\rho(T)]/dT$\ we can identify for the UD sample a structural transition at $T_s = (62 \pm 10)$~K from the minimum of the $\rho(T)$\ curve, and finally a sharp superconducting transition onset at $T_c = (25 \pm 0.2)$~K (Fig. 2).
 
 \begin{table}
 \centering
  \caption{Superconducting transition parameters for $\rm Ba(Fe_{1-x}Co_x)_2As_2$\ single crystals.}\label{tab1}
  \medskip
\begin{tabular}{|c|c|c|c|c|}
\hline
 Sample & Doping level, $x$ & $T_c^{on}$,~K & $T_c^m$,~K & $\Delta T_c$,~K \\
\hline
 UD & 0.07 & 26 & 23.8 & 5.2 \\
 \hline
 OP & 0.09 & 26.9 & 25.9 & 1.4 \\
 \hline
 OD & 0.11 & 23.7 & 21.5 & 3.5 \\
 \hline
 \end{tabular}
 \end{table}

The MMWA measurements on these samples were carried out in the temperature range of $10\div 300$~K and the magnetic field range of $0\div 0.75$~T. The experiments were performed on a Bruker BER-418s ESR-spectrometer at a frequency of 9.4~GHz (X-band). The DC field was modulated with a frequency of 100~kHz with the amplitude of 0.1 to 10~Oe. The MMWA signal is detected and amplified by the lock-in amplifier at the first harmonic of the modulation. The crystal $ab$-plane was oriented perpendicular to the applied magnetic field. The MMWA signal was registered at fixed temperature upon sweeping the magnetic field up and down. We have observed the field dependent hysteresis of the MMWA signal which reflects the magnetic irreversibility due to vortex pinning (see Fig. 3). The hysteresis amplitude $L$\ was determined from experimental MMWA loops as the difference between magnitudes of the power absorbed upon sweeping the magnetic field up ($P_{up}$) and down ($P_{down}$), $L = P_{up} - P_{down}$.  

\begin{figure}[t]
\includegraphics[width=7.5cm]{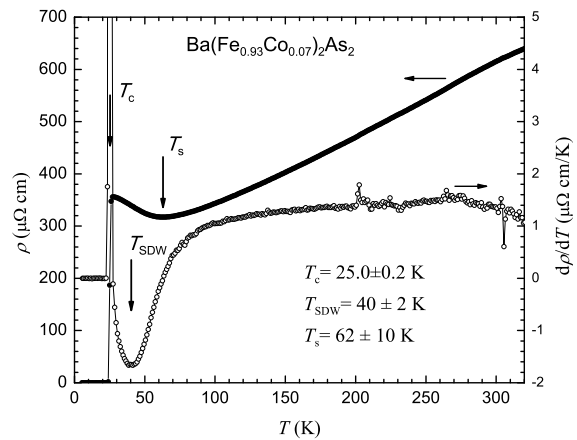}
\caption{\label{rho} Temperature dependence of resistivity (closed circles) and its derivative $d\rho /dT$\ (open circles) for the UD sample, revealing the structural $T_s$, magnetic $T_{SDW}$\ and superconducting $T_c$\ transitions.   }
\end{figure}

The measurement protocol was as follows: A sample was cooled from $T > T_c$ down to the measurement temperature in a fixed magnetic field $\mu_0H_{res} = 5$~mT (residual field of the spectrometer magnet). Then the applied DC field was swept with a rate of 4~mT/s from $H_{res}$\ to $H_{max}$\ and back. $\mu_0H_{max}$\ was varied from 0.5 to 0.75~T depending on a particular measurement. Further details on the measurement technique, procedure and the physical basis of the MMWA method can be found in Refs.\cite{Shaposhnikova98,Panarina10}. The hysteresis amplitude $L$\ appears to be a function of cobalt concentration, magnetic field, and temperature.

\section{Results and discussion}

In the temperature range near to the critical temperature the hysteresis loop closes at a certain field value \hir called the irreversibility field. The temperature dependence of the irreversibility field $H_{irr}(T)$\ defines the irreversibility line, which separates the areas with zero and nonzero values of the superconducting critical current density on the temperature/magnetic field phase diagram. The plot of \hir versus the reduced temperature $T/T_c$\ (Fig. 4) allows correct comparison of the irreversibility lines of the samples with different critical temperature. One can see that the irreversibility line of the sample with doping level $x = 0.07$\ (UD) is the steepest. 

The temperature dependence of the amplitude of the MMWA hysteresis loop $L(T)$\ was obtained at fixed values of the applied magnetic field. The $L(T)$\ dependence for three \ba single crystals is presented in Fig. 5 for two selected fields $\mu_0H_a = 0.1$\ and 0.45~T. These fields are indicated by dotted vertical lines in Fig. 3. The field 0.1~T was chosen to be somewhat higher than the position of the low-field peak occurring at $\mu_0H_a < 0.05$~T. This peak arises due to two competing effects. The increase of the MMWA signal upon starting the sweep of the applied field is due to the rearrangement of the vortex distribution. When the field begins to rise, the superconducting sample is gradually filled by vortices, entering the sample from edges towards the center. In this regime the growth of the MMWA signal is due to the increase of the number of vortices [the term $F_{\uparrow}$ in Eq.(1) in fact describes this increase, see below]. As soon as the first vortex reaches the center of the sample, the volume through which the critical current flows does not change anymore. The magnitude of the MMWA signal is now determined by the critical current density which decreases with increasing the field strength. Therefore the MMWA signal decreases. These two effects are responsible for a characteristic peak of the MMWA response at low fields (Fig. 3). The second (higher) magnetic field value of 0.45~T was chosen such, so that to collect a sufficient number of experimental data points in the irreversible area where $L \neq 0$.

\begin{figure}
\includegraphics[width=8cm]{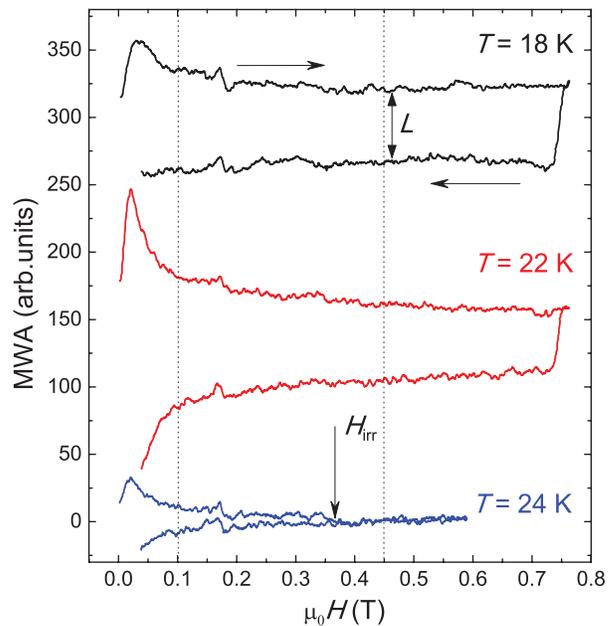}
\caption{\label{loops} MMWA hysteresis loops for the UD $\rm Ba(Fe_{1-x}Co_x)_2As_2$\ single crystal with $x = 0.07$\ at temperatures 18~K, 22~K, and 24~K (from top to bottom). The loops are shifted on the vertical scale relative to each other for convenience. Horizontal arrows show the direction of the magnetic field sweep (first up-field followed by down-field). The vertical arrow indicates the irreversibility field at which the hysteresis loop closes. The vertical dotted lines indicate the fields at which the amplitude of the loop $L$\ was measured as a function of temperature. A small feature at $ \sim 0.175$~T corresponds to an EPR signal from iron ions present in the glass ampoule in which the sample was placed. } 
\end{figure} 

\begin{figure}
\includegraphics[width=7.5cm]{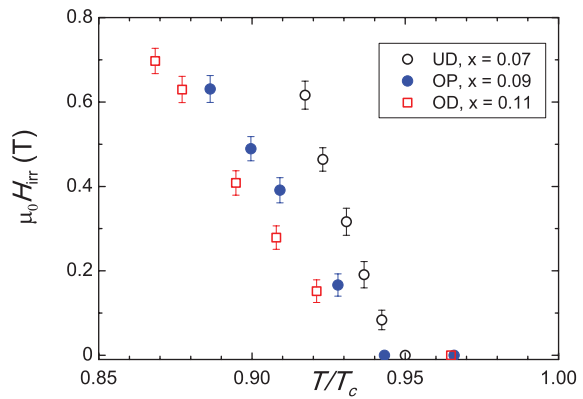}
\caption{\label{IL} Irreversibility field versus reduced temperature for three samples of $\rm Ba(Fe_{1-x}Co_x)_2As_2$. }
\end{figure}

For the UD sample the $L$\ first increases with decreasing temperature, reaches a maximum at $T = 21.5$~K and then decreases. The position of this maximum is independent of the magnetic field value. The $L(T)$\ dependence for the OD sample is different: Near the superconducting transition the amplitude of the loop increases quite steeply and tends to level off in the lower temperature range. The OP sample reveals a behavior intermediate between UD and OD. At $\mu_0H_a = 0.1$~T the $L(T)$\ dependence is similar to that of the UD sample, whereas at a higher field $\mu_0H_a = 0.45$~T it is similar to that of the OD sample.

The analytical form of $L(T)$\ is a complex function of the temperature dependent critical current \jc, the amplitude $u$\ of thermal fluctuations of the flux-line positions and the viscosity of the vortex matter $\eta$\ \cite{Shaposhnikova98}:

\begin{eqnarray}
 L(T,H) = &&C_0 \frac{j_c^2(T,H)}{\eta^{7/2}(T)}\exp\bigg(-8 \pi^2 \frac{\langle u^2(T) \rangle }{a^2}\bigg)\nonumber\\
   					&&\times\left[ F_{\uparrow}(H,j) - F_{\downarrow}(H,j)\right] 
 \end{eqnarray}
  Here $C_0$\ is a constant defined by the sample size and by the registration conditions of the hysteresis loop, and $a$\ is the mean distance between pinning centers. Function (1) includes integrals over the instantaneous distributions of field and current in a sample for sweeping the field up ($F_{\uparrow}(H,j)$) and down ($F_{\downarrow}(H,j)$). The distributions depend on the shape and sizes of the sample, as well as on the critical current density. A detailed description of the $L(T,H)$\ function and its components is given in Ref. 10, where a theoretical model of the hysteresis of the microwave absorption has been developed.

The $\eta(T)$\ dependence can be obtained from the relation $\eta = \Phi_0B_{c2}/\rho_n$ , where $\Phi_0$\ is the magnetic flux quantum, $\rho_n$\ is the normal-state resistivity \cite{Bardeen65}. The temperature dependence of the upper critical field $B_{c2}(T)$\ can be  derived from the Ginsburg-Landau theory. Then one gets

\begin{equation}
	\eta = \eta_0 \frac{1-(T/T_c)^2}{1+(T/T_c)^2}
\end{equation}

with $\eta_0\equiv\eta(T=0)$.

Eq. (1) depends exponentially on the ratio of the mean square of thermal fluctuations $\langle u^2 \rangle$\ to the square of the mean distance between pinning centers $a^2$. According to Ref.\cite{Feigel'man90} the fluctuation amplitude depends on temperature as follows from the equation $\langle u^2 \rangle = 4\pi T\lambda^2/\Phi_0^{3/2}B^{1/2}$. The general formula of the temperature dependence of the field penetration depth has the form $\lambda(T)=\lambda_0(1-(T/T_c)^m)^{-1/2}$. For \ba the index $m$\ varies from 2 for the UD samples to $\sim 2.5$\ for the OP and OD compounds \cite{Gordon09,Prozorov09c,Gordon10}.  Thus, the temperature dependence of the ratio can be written as

\begin{equation}
	\frac{\langle u^2(T) \rangle }{a^2} = c' \frac{T}{1-(T/T_c)^m}
\end{equation}

\noindent Here $c'$\ is the parameter that depends on the material properties, in particular on the form and on the number of the structural defects and impurities, which serve as pinning centers. It also depends on the magnetic field, specifically at low fields usually $c' \propto B^{-1/2}$. 

There is a number of factors that affect the $j_c(T)$\ function, such as the chemical composition, the material state (e.g., poly- or  single crystals, or thin films), type and concentration of defects, magnetic field strength etc. \cite{Blatter94}. It was found experimentally \cite{Prozorov08} that in \ba single crystals the critical current density is a power function of temperature which can be written as:  
 
\begin{equation}
	j_c = j_{c0} (1-(T/T_c)^p)^n  .
\end{equation}

\noindent Here $j_{c0}$\ is the critical current density at zero temperature, and $p$\ and $n$\ are indices defined by the type of pinning centers. This form of the power law extensively discussed in Refs.\cite{Schnack93,Griessen94,Ijaduola06}  is characteristic for the most of high-$T_c$\ superconductors including cuprates. According to theoretical studies \cite{Schnack93,Griessen94}, in low magnetic field these indexes amount to $p = 2$\ and $n = 5/2$\ for the \dl-pinning inherent in superconductors with weak pinning centers of a point type. (The \dl-pinning arises due to spatial variations in the charge carrier mean free path $l$\ and local lowering of the superconducting order parameter near lattice defects \cite{Blatter94}.) The presence of extended defects (in particular, twin domain boundaries or impurity phase inclusions) results in the so-called \dtc-pinning with $p = 2$\ and $n = 7/6$. (\dtc-pinning is associated with disorder at critical temperature \tc.)

\begin{figure}
\includegraphics[width=7.5cm]{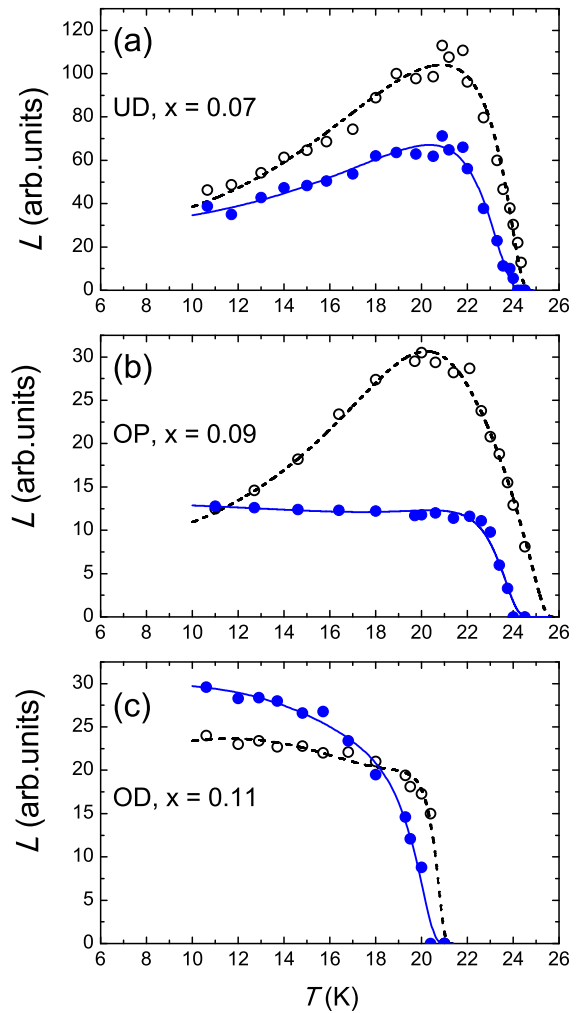}
\caption{\label{Tdep} Temperature dependences of the MMWA hysteresis loop amplitude for the UD (a), OP (b), and OD single crystals (c) at a magnetic field of 0.1 T (open circles) and 0.45 T (closed circles), respectively.  Solid and dashed curves are calculated with Eq. (1). }
\end{figure}

We simulated the temperature dependence of the hysteresis amplitude using Eq.(1) and taking the temperature functions $\eta(T)$, $\langle u^2(T) \rangle$\ and $j_c(T)$\ given by Eqs.(2-4) into account.  The only fitting parameters were $j_{c0}$, $c'$, $n$\ and $C'_0=C_0/\eta_0^{7/2}$. The index $p$\ was fixed at $p = 2$\ in accordance with the theory requirements \cite{Blatter94}. We used $m = 2$\ for the UD sample and $m = 2.5$\ for description of the data for OP and OD samples in accordance with Refs.\cite{Gordon09,Prozorov09c,Gordon10}. The results of the simulation are shown in Fig. 5 by solid and dashed lines for the applied field of 0.45~T and 0.1~T, respectively. The obtained fitting parameters are presented in Table 2. (The parameter $C'_0$  is not given there as it consists of several cofactors which cannot be separated.)

The $j_{c0}$\ parameter defines the critical current density. First of all, it should be noted that its magnitude is in a good agreement with the data obtained from transport and magnetic measurements \cite{Prozorov08,Yamamoto09}. For the UD sample it has the temperature dependence described by Eq.(4) with the $n$\ index close to 1.2 which corresponds to the \dtc-pinning. According to theoretical predictions \cite{Blatter94,Griessen94} and experimental findings \cite{Griessen94,Ijaduola06}, the \dtc-pinning occurs in a superconductor with large-size defects. In particular, such $j_c(T)$\ dependence is observed in YBCO films containing large-size ($\sim 30$~nm) inclusions of the impurity phase, which provide effective vortex pinning \cite{Ijaduola06}. Thus, taking into account that in the UD crystal the value of $n$\ is close to that of the \dtc-pinning regime, a large $j_{c0}$\ value and its weak field dependence can be related to the occurrence of large-size pinning centers in this sample. Considering the phase diagram of the \ba compound \cite{Aswartham11,Wang09} where the coexistence of the superconducting and magnetically ordered phases at $x\leq 0.8$\ has been observed \cite{Lester09,Pratt09}, one can associate such large-size pinning centers with regions of magnetically ordered phase which should apparently be present in the UD single crystal below $T_{SDW} = 40$~K. In favor of this assumption, the study of a similar iron pnictide $\rm Ba_{1-x}K_xFe_2As_2$\ by neutron scattering, muon-spin rotation and magnetic force microscopy revealed magnetic (nonsuperconducting) inclusions of the 65~nm size \cite{Park09}. We conjecture that magnetically ordered domains in our UD crystal have the size of the same order of magnitude.

\begin{table}
 \centering
  \caption{Parameters $j_{c0}$, $n$\ and $c'$\ obtained by fitting Eq.(1) to the experimental dependence $L(T)$ at two magnitudes of the applied magnetic field \ha.}\label{tab2}
  \medskip
\begin{tabular}{|c|c|c|c|c|}
\hline
 Sample & $\mu_0H_a$, T & $j_{c0}$,~A/cm$^2$ & $n$ & $c'$,~1/K \\
\hline
 UD & 0.1 & $1.7\cdot 10^4$ & $1.3\pm0.2$ & $(4.5\pm0.5)\cdot10^{-4}$ \\
 \hline
 UD & 0.45 & $1.5\cdot 10^4$ & $1.1\pm0.2$ & $(5.8\pm0.7)\cdot10^{-4}$ \\
 \hline
 OP & 0.1 & $0.95\cdot 10^4$ & $1.2\pm0.2$ & $(6.8\pm0.8)\cdot10^{-4}$ \\
 \hline
 OP & 0.45 & $0.88\cdot 10^4$ & $2.6\pm0.4$ & $(4.6\pm1.3)\cdot10^{-4}$ \\
 \hline
 OD & 0.1 & $1.4\cdot 10^4$ & $2.9\pm0.5$ & $(5.5\pm1.2)\cdot10^{-4}$ \\
 \hline
 OD & 0.45 & $1.9\cdot 10^4$ & $1.9\pm0.5$ & $(9.8\pm2.4)\cdot10^{-4}$ \\
 \hline
 \end{tabular}
 \end{table}

Another origin of the large-scale defects and their influence on the critical current density was discussed in Ref.\cite{Prozorov09b}. Twin boundaries were considered as such extended defects in \ba crystals with $x \leq 0.058$\ grown by self-flux technique. However, in crystals with higher Co concentration no twin boundary structure was revealed. For our UD sample with $x=0.07$\ we therefore do not expect twin boundaries to occur. 

The decrease of $j_{c0}$\ and the change of the $n$\ index towards the value of 2.5 corresponding to the \dl-pinning (at least in the high-field range) in the OP sample (see Table 2) is indicative of the pinning due to weak point defects\cite{comment}. As suggested in Ref.\cite{Beek10} such pinning centers can be formed as a result of inhomogeneous distribution of the dopant atoms. It is therefore likely that also in our OP and OD samples the Co impurity atoms serve as point pinning centers. This assumption is supported by the observation of the increase of the critical current $j_{c0}$\ and of the $c'$\ parameter, related to the center density, upon transition from OP to OD conditions. This can be naturally related with the increasing number of pinning centers. Some enhancement of the $j_{c0}$\ value of the OD crystal with the applied magnetic field is probably connected with the fish-tail effect observed frequently in \ba  (see, for example, Refs.\cite{Prozorov08,Yamamoto09}). We note however that the steepest irreversibility line is observed for the UD \ba single crystal (Fig. 4) suggesting that magnetically ordered phase inclusions are most effective pinning centers for superconducting vortices. 

Considering the data in Table 2 one should note that the parameter $c'$\ does not obey the $B^{-1/2}$\ dependence, because it is valid in the low field regime only. The $c'$\  even increases with $B$\ for the UD and OD samples. In the latter case the increase is most pronounced. For this sample also the $j_c$\ increases which may be related to the fish-tail effect, as discussed above. The functional dependence $<u^2>\propto  B^{-1/2}$\ is a consequence of the dependence of $<u^2>$\ on the shear modulus $c_{66}$, which in turn is proportional to $B$ \cite{Feigel'man90, Blatter94}. This relationship holds well in the low field regime. However, with increasing the field two additional effects may become relevant. The first effect is the softening of the shear modulus which yields a decrease of $c_{66}$, as discussed in Ref.\cite{Pippard69}. The second effect pointed out in Ref.\cite{Larkin79} is the formation of vortex bundles. Both effects can result in the increase of the vortex shift $\sqrt{<u^2>}$\ under the action of thermal fluctuations, as compared with the rigid (elastic) vortex lattice. Plausibly these effects may be responsible for an increase of $c'$\ and also of $j_c$\ with magnetic field in the case of the OD sample.

It is possible to estimate the mean value of the thermal fluctuation amplitude  $u=\sqrt{<u^2>}$\  and the distance between pinning centers $a$\ on the basis of the obtained data and the known functional dependences. Assuming that the above discussed $<u^2(B)>$\ dependence is valid at low fields and taking the value $\lambda_0=2080$ \AA\cite{Gordon09b}, we obtain the following estimates of $u$\  and $a$\ for the UD, OP and OD samples at $T=15$~K and $H=0.1$~T, respectively:  $u  = 27.7$~\AA, 25~\AA, 28~\AA; and $a=$ 263~\AA, 214~\AA, 236~\AA. One can see that the maximum distance between pinning centers, which corresponds to the minimum of their density, is revealed for the UD sample. This is in agreement with the minimum doping level in this sample and supports an assumption on the droplet form of pinning centers. A somewhat surprising increase of $a$\ in the OD sample, as compared with the OP sample, suggests a possible clustering of dopant atoms. In this scenario the concentration of clusters is smaller than that of point defects. However they are more efficient as pinning centers yielding an increase of the critical current density (see Table 2).

\section{Conclusions}

A set of the \ba single crystals with Co concentration $x=0.07$, 0.09, and 0.11 was studied by the MMWA method. The analysis of the temperature and field dependences of the MMWA hysteresis amplitude enables to estimate the critical current density and the critical indices of the $j_c(T)$\ function, and to determine their dependence on temperature and Co doping. On the basis of our data we conclude that in the underdoped crystal ($x = 0.07$) the pinning is of a \dtc\ type related to the presence of regions of a nonsuperconducting (magnetically ordered) phase which provides the steepest slope of the irreversibility line for the underdoped sample. On the other hand, in the optimally doped and overdoped single crystals the pinning is caused by structural imperfections due to an inhomogeneous distribution of cobalt dopant. Because of this, the \dl-pinning is inherent in these crystals.

\section*{Acknowledgments}

This work has been supported by the Russian Foundation for Basic Research (RFFI) under Grant no. 10-02-01056, by the Deutsche Forschungsgemeinschaft (DFG) through the priority program SPP 1458 (Grants BE 1744/13 and GR 3330/2), the DFG Graduate School GRK 1621 and  the European Commission through the LOTHERM project (PITN-GA-2009-238475). A German Russian collaborative research grant of the DFG  BU 887/13-2 and of the RFFI 12-02-91339-NNIO  is gratefully acknowledged. SW acknowledges funding by the DFG through project WU 595/3-1.   



\begin{thebibliography}{27}
\expandafter\ifx\csname natexlab\endcsname\relax\def\natexlab#1{#1}\fi
\expandafter\ifx\csname bibnamefont\endcsname\relax
  \def\bibnamefont#1{#1}\fi
\expandafter\ifx\csname bibfnamefont\endcsname\relax
  \def\bibfnamefont#1{#1}\fi
\expandafter\ifx\csname citenamefont\endcsname\relax
  \def\citenamefont#1{#1}\fi
\expandafter\ifx\csname url\endcsname\relax
  \def\url#1{\texttt{#1}}\fi
\expandafter\ifx\csname urlprefix\endcsname\relax\def\urlprefix{URL }\fi
\providecommand{\bibinfo}[2]{#2}
\providecommand{\eprint}[2][]{\url{#2}}

\bibitem[{\citenamefont{Aswartham et~al.}(2011)\citenamefont{Aswartham, Nacke,
  Friemel, Leps, Wurmehl, Wizent, Hess, Klingeler, Behr, Singh
  et~al.}}]{Aswartham11}
\bibinfo{author}{\bibfnamefont{S.}~\bibnamefont{Aswartham}},
  \bibinfo{author}{\bibfnamefont{C.}~\bibnamefont{Nacke}},
  \bibinfo{author}{\bibfnamefont{G.}~\bibnamefont{Friemel}},
  \bibinfo{author}{\bibfnamefont{N.}~\bibnamefont{Leps}},
  \bibinfo{author}{\bibfnamefont{S.}~\bibnamefont{Wurmehl}},
  \bibinfo{author}{\bibfnamefont{N.}~\bibnamefont{Wizent}},
  \bibinfo{author}{\bibfnamefont{C.}~\bibnamefont{Hess}},
  \bibinfo{author}{\bibfnamefont{R.}~\bibnamefont{Klingeler}},
  \bibinfo{author}{\bibfnamefont{G.}~\bibnamefont{Behr}},
  \bibinfo{author}{\bibfnamefont{S.}~\bibnamefont{Singh}},
  \bibnamefont{et~al.}, \bibinfo{journal}{J. Cryst. Growth}
  \textbf{\bibinfo{volume}{314}}, \bibinfo{pages}{341} (\bibinfo{year}{2011}).

\bibitem[{\citenamefont{Prozorov et~al.}(2008)\citenamefont{Prozorov, Ni,
  Tanatar, Kogan, Gordon, Martin, Blomberg, Prommapan, Yan, Bud'ko
  et~al.}}]{Prozorov08}
\bibinfo{author}{\bibfnamefont{R.}~\bibnamefont{Prozorov}},
  \bibinfo{author}{\bibfnamefont{N.}~\bibnamefont{Ni}},
  \bibinfo{author}{\bibfnamefont{M.~A.} \bibnamefont{Tanatar}},
  \bibinfo{author}{\bibfnamefont{V.~G.} \bibnamefont{Kogan}},
  \bibinfo{author}{\bibfnamefont{R.~T.} \bibnamefont{Gordon}},
  \bibinfo{author}{\bibfnamefont{C.}~\bibnamefont{Martin}},
  \bibinfo{author}{\bibfnamefont{E.~C.} \bibnamefont{Blomberg}},
  \bibinfo{author}{\bibfnamefont{P.}~\bibnamefont{Prommapan}},
  \bibinfo{author}{\bibfnamefont{J.~Q.} \bibnamefont{Yan}},
  \bibinfo{author}{\bibfnamefont{S.~L.} \bibnamefont{Bud'ko}},
  \bibnamefont{et~al.}, \bibinfo{journal}{Phys.\ Rev.\ B}
  \textbf{\bibinfo{volume}{78}}, \bibinfo{pages}{224506}
  (\bibinfo{year}{2008}).

\bibitem[{\citenamefont{Prozorov
  et~al.}(2009{\natexlab{a}})\citenamefont{Prozorov, Tanatar, Blomberg,
  Prommapan, Gordon, Ni, Bud'ko, and Canfield}}]{Prozorov09}
\bibinfo{author}{\bibfnamefont{R.}~\bibnamefont{Prozorov}},
  \bibinfo{author}{\bibfnamefont{M.~A.} \bibnamefont{Tanatar}},
  \bibinfo{author}{\bibfnamefont{E.~C.} \bibnamefont{Blomberg}},
  \bibinfo{author}{\bibfnamefont{P.}~\bibnamefont{Prommapan}},
  \bibinfo{author}{\bibfnamefont{R.~T.} \bibnamefont{Gordon}},
  \bibinfo{author}{\bibfnamefont{N.}~\bibnamefont{Ni}},
  \bibinfo{author}{\bibfnamefont{S.~L.} \bibnamefont{Bud'ko}},
  \bibnamefont{and} \bibinfo{author}{\bibfnamefont{P.~C.}
  \bibnamefont{Canfield}}, \bibinfo{journal}{Physica C}
  \textbf{\bibinfo{volume}{469}}, \bibinfo{pages}{667}
  (\bibinfo{year}{2009}{\natexlab{a}}).

\bibitem[{\citenamefont{Yamamoto et~al.}(2009)\citenamefont{Yamamoto,
  Jaroszynski, Tarantini, Balicas, Jiang, Gurevich, Larbalestier, Jin, Sefat,
  McGuire et~al.}}]{Yamamoto09}
\bibinfo{author}{\bibfnamefont{A.}~\bibnamefont{Yamamoto}},
  \bibinfo{author}{\bibfnamefont{J.}~\bibnamefont{Jaroszynski}},
  \bibinfo{author}{\bibfnamefont{C.}~\bibnamefont{Tarantini}},
  \bibinfo{author}{\bibfnamefont{L.}~\bibnamefont{Balicas}},
  \bibinfo{author}{\bibfnamefont{J.}~\bibnamefont{Jiang}},
  \bibinfo{author}{\bibfnamefont{A.}~\bibnamefont{Gurevich}},
  \bibinfo{author}{\bibfnamefont{D.}~\bibnamefont{Larbalestier}},
  \bibinfo{author}{\bibfnamefont{R.}~\bibnamefont{Jin}},
  \bibinfo{author}{\bibfnamefont{A.}~\bibnamefont{Sefat}},
  \bibinfo{author}{\bibfnamefont{M.}~\bibnamefont{McGuire}},
  \bibnamefont{et~al.}, \bibinfo{journal}{Appl.\ Phys.\ Lett.}
  \textbf{\bibinfo{volume}{94}}, \bibinfo{pages}{062511}
  (\bibinfo{year}{2009}).

\bibitem[{\citenamefont{Prozorov
  et~al.}(2009{\natexlab{b}})\citenamefont{Prozorov, Tanatar, Ni, Kreyssig,
  Nandi, Bud'ko, Goldman, and Canfield}}]{Prozorov09b}
\bibinfo{author}{\bibfnamefont{R.}~\bibnamefont{Prozorov}},
  \bibinfo{author}{\bibfnamefont{M.~A.} \bibnamefont{Tanatar}},
  \bibinfo{author}{\bibfnamefont{N.}~\bibnamefont{Ni}},
  \bibinfo{author}{\bibfnamefont{A.}~\bibnamefont{Kreyssig}},
  \bibinfo{author}{\bibfnamefont{S.}~\bibnamefont{Nandi}},
  \bibinfo{author}{\bibfnamefont{S.}~\bibnamefont{Bud'ko}},
  \bibinfo{author}{\bibfnamefont{A.~I.} \bibnamefont{Goldman}},
  \bibnamefont{and} \bibinfo{author}{\bibfnamefont{P.~C.}
  \bibnamefont{Canfield}}, \bibinfo{journal}{Phys.\ Rev.\ B}
  \textbf{\bibinfo{volume}{80}}, \bibinfo{pages}{174517}
  (\bibinfo{year}{2009}{\natexlab{b}}).

\bibitem[{\citenamefont{Lester et~al.}(2009)\citenamefont{Lester, Chu,
  Analytis, Capelli, Erickson, Condron, Toney, Fisher, and Hayden}}]{Lester09}
\bibinfo{author}{\bibfnamefont{C.}~\bibnamefont{Lester}},
  \bibinfo{author}{\bibfnamefont{J.-H.} \bibnamefont{Chu}},
  \bibinfo{author}{\bibfnamefont{J.~G.} \bibnamefont{Analytis}},
  \bibinfo{author}{\bibfnamefont{S.~C.} \bibnamefont{Capelli}},
  \bibinfo{author}{\bibfnamefont{A.~S.} \bibnamefont{Erickson}},
  \bibinfo{author}{\bibfnamefont{C.~L.} \bibnamefont{Condron}},
  \bibinfo{author}{\bibfnamefont{M.~F.} \bibnamefont{Toney}},
  \bibinfo{author}{\bibfnamefont{I.~R.} \bibnamefont{Fisher}},
  \bibnamefont{and} \bibinfo{author}{\bibfnamefont{S.~M.}
  \bibnamefont{Hayden}}, \bibinfo{journal}{Phys.\ Rev.\ B}
  \textbf{\bibinfo{volume}{79}}, \bibinfo{pages}{144523}
  (\bibinfo{year}{2009}).

\bibitem[{\citenamefont{Chu et~al.}(2009)\citenamefont{Chu, Analytis,
  Kucharczyk, and Fisher}}]{Chu09}
\bibinfo{author}{\bibfnamefont{J.-H.} \bibnamefont{Chu}},
  \bibinfo{author}{\bibfnamefont{J.~G.} \bibnamefont{Analytis}},
  \bibinfo{author}{\bibfnamefont{C.}~\bibnamefont{Kucharczyk}},
  \bibnamefont{and} \bibinfo{author}{\bibfnamefont{I.~R.}
  \bibnamefont{Fisher}}, \bibinfo{journal}{Phys.\ Rev.\ B}
  \textbf{\bibinfo{volume}{79}}, \bibinfo{pages}{014506}
  (\bibinfo{year}{2009}).

\bibitem[{\citenamefont{Ning et~al.}(2009)\citenamefont{Ning, Ahilan, Imai,
  Sefat, Jin, Mcguire, Sales, and Mandrus}}]{Ning09}
\bibinfo{author}{\bibfnamefont{F.}~\bibnamefont{Ning}},
  \bibinfo{author}{\bibfnamefont{K.}~\bibnamefont{Ahilan}},
  \bibinfo{author}{\bibfnamefont{T.}~\bibnamefont{Imai}},
  \bibinfo{author}{\bibfnamefont{A.~S.} \bibnamefont{Sefat}},
  \bibinfo{author}{\bibfnamefont{R.}~\bibnamefont{Jin}},
  \bibinfo{author}{\bibfnamefont{M.~A.} \bibnamefont{Mcguire}},
  \bibinfo{author}{\bibfnamefont{B.~C.} \bibnamefont{Sales}}, \bibnamefont{and}
  \bibinfo{author}{\bibfnamefont{D.}~\bibnamefont{Mandrus}},
  \bibinfo{journal}{J.\ Phys.\ Soc.\ Jpn.} \textbf{\bibinfo{volume}{78}},
  \bibinfo{pages}{013711} (\bibinfo{year}{2009}).

\bibitem[{\citenamefont{Wang et~al.}(2009)\citenamefont{Wang, Wu, Wu, Liu,
  Chen, Xie, and Chen}}]{Wang09}
\bibinfo{author}{\bibfnamefont{X.~F.} \bibnamefont{Wang}},
  \bibinfo{author}{\bibfnamefont{T.}~\bibnamefont{Wu}},
  \bibinfo{author}{\bibfnamefont{G.}~\bibnamefont{Wu}},
  \bibinfo{author}{\bibfnamefont{R.~H.} \bibnamefont{Liu}},
  \bibinfo{author}{\bibfnamefont{H.}~\bibnamefont{Chen}},
  \bibinfo{author}{\bibfnamefont{Y.~L.} \bibnamefont{Xie}}, \bibnamefont{and}
  \bibinfo{author}{\bibfnamefont{X.~H.} \bibnamefont{Chen}},
  \bibinfo{journal}{New\ J.\ Phys.} \textbf{\bibinfo{volume}{11}},
  \bibinfo{pages}{045003} (\bibinfo{year}{2009}).

\bibitem[{\citenamefont{Shaposhnikova et~al.}(1998)\citenamefont{Shaposhnikova,
  Vashakidze, Khasanov, and Talanov}}]{Shaposhnikova98}
\bibinfo{author}{\bibfnamefont{T.}~\bibnamefont{Shaposhnikova}},
  \bibinfo{author}{\bibfnamefont{Y.}~\bibnamefont{Vashakidze}},
  \bibinfo{author}{\bibfnamefont{R.}~\bibnamefont{Khasanov}}, \bibnamefont{and}
  \bibinfo{author}{\bibfnamefont{Y.}~\bibnamefont{Talanov}},
  \bibinfo{journal}{Physica C} \textbf{\bibinfo{volume}{30}}
  (\bibinfo{year}{1998}).

\bibitem[{\citenamefont{Panarina et~al.}(2010)\citenamefont{Panarina, Talanov,
  Shaposhnikova, Beysengulov, Vavilova, G.Behr, Kondrat, Hess, Leps, Wurmehl
  et~al.}}]{Panarina10}
\bibinfo{author}{\bibfnamefont{N.~Y.} \bibnamefont{Panarina}},
  \bibinfo{author}{\bibfnamefont{Y.~I.} \bibnamefont{Talanov}},
  \bibinfo{author}{\bibfnamefont{T.~S.} \bibnamefont{Shaposhnikova}},
  \bibinfo{author}{\bibfnamefont{N.~R.} \bibnamefont{Beysengulov}},
  \bibinfo{author}{\bibfnamefont{E.}~\bibnamefont{Vavilova}},
  \bibinfo{author}{\bibnamefont{G.Behr}},
  \bibinfo{author}{\bibfnamefont{A.}~\bibnamefont{Kondrat}},
  \bibinfo{author}{\bibfnamefont{C.}~\bibnamefont{Hess}},
  \bibinfo{author}{\bibfnamefont{N.}~\bibnamefont{Leps}},
  \bibinfo{author}{\bibfnamefont{S.}~\bibnamefont{Wurmehl}},
  \bibnamefont{et~al.}, \bibinfo{journal}{Phys.\ Rev.\ B}
  \textbf{\bibinfo{volume}{81}}, \bibinfo{pages}{224509}
  (\bibinfo{year}{2010}).

\bibitem[{\citenamefont{Bardeen and Stephen}(1965)}]{Bardeen65}
\bibinfo{author}{\bibfnamefont{J.}~\bibnamefont{Bardeen}} \bibnamefont{and}
  \bibinfo{author}{\bibfnamefont{M.~J.} \bibnamefont{Stephen}},
  \bibinfo{journal}{Phys.\ Rev.} \textbf{\bibinfo{volume}{140}},
  \bibinfo{pages}{A1197} (\bibinfo{year}{1965}).

\bibitem[{\citenamefont{Feigel'man and Vinokur}(1990)}]{Feigel'man90}
\bibinfo{author}{\bibfnamefont{M.~V.} \bibnamefont{Feigel'man}}
  \bibnamefont{and} \bibinfo{author}{\bibfnamefont{V.~M.}
  \bibnamefont{Vinokur}}, \bibinfo{journal}{Phys.\ Rev.\ B}
  \textbf{\bibinfo{volume}{41}}, \bibinfo{pages}{8986} (\bibinfo{year}{1990}).

\bibitem[{\citenamefont{Gordon et~al.}(2009{\natexlab{a}})\citenamefont{Gordon,
  Martin, Kim, Ni, Tanatar, Schmalian, Mazin, Bud'ko, Canfield, and
  Prozorov}}]{Gordon09}
\bibinfo{author}{\bibfnamefont{R.~T.} \bibnamefont{Gordon}},
  \bibinfo{author}{\bibfnamefont{C.}~\bibnamefont{Martin}},
  \bibinfo{author}{\bibfnamefont{H.}~\bibnamefont{Kim}},
  \bibinfo{author}{\bibfnamefont{N.}~\bibnamefont{Ni}},
  \bibinfo{author}{\bibfnamefont{M.~A.} \bibnamefont{Tanatar}},
  \bibinfo{author}{\bibfnamefont{J.}~\bibnamefont{Schmalian}},
  \bibinfo{author}{\bibfnamefont{I.~I.} \bibnamefont{Mazin}},
  \bibinfo{author}{\bibfnamefont{S.~L.} \bibnamefont{Bud'ko}},
  \bibinfo{author}{\bibfnamefont{P.~C.} \bibnamefont{Canfield}},
  \bibnamefont{and} \bibinfo{author}{\bibfnamefont{R.}~\bibnamefont{Prozorov}},
  \bibinfo{journal}{Phys.\ Rev.\ B} \textbf{\bibinfo{volume}{79}},
  \bibinfo{pages}{100506R} (\bibinfo{year}{2009}{\natexlab{a}}).

\bibitem[{\citenamefont{Prozorov
  et~al.}(2009{\natexlab{c}})\citenamefont{Prozorov, Tanatar, Gordon, Martin,
  Kim, Kogan, Ni, Tillman, Bud'ko, and Canfield}}]{Prozorov09c}
\bibinfo{author}{\bibfnamefont{R.}~\bibnamefont{Prozorov}},
  \bibinfo{author}{\bibfnamefont{M.~A.} \bibnamefont{Tanatar}},
  \bibinfo{author}{\bibfnamefont{R.~T.} \bibnamefont{Gordon}},
  \bibinfo{author}{\bibfnamefont{C.}~\bibnamefont{Martin}},
  \bibinfo{author}{\bibfnamefont{H.}~\bibnamefont{Kim}},
  \bibinfo{author}{\bibfnamefont{V.~G.} \bibnamefont{Kogan}},
  \bibinfo{author}{\bibfnamefont{N.}~\bibnamefont{Ni}},
  \bibinfo{author}{\bibfnamefont{M.~E.} \bibnamefont{Tillman}},
  \bibinfo{author}{\bibfnamefont{S.~L.} \bibnamefont{Bud'ko}},
  \bibnamefont{and} \bibinfo{author}{\bibfnamefont{P.~C.}
  \bibnamefont{Canfield}}, \bibinfo{journal}{Physica C}
  \textbf{\bibinfo{volume}{469}}, \bibinfo{pages}{582}
  (\bibinfo{year}{2009}{\natexlab{c}}).

\bibitem[{\citenamefont{Gordon et~al.}(2010)\citenamefont{Gordon, Kim, Tanatar,
  Prozorov, and Kogan}}]{Gordon10}
\bibinfo{author}{\bibfnamefont{R.~T.} \bibnamefont{Gordon}},
  \bibinfo{author}{\bibfnamefont{H.}~\bibnamefont{Kim}},
  \bibinfo{author}{\bibfnamefont{M.~A.} \bibnamefont{Tanatar}},
  \bibinfo{author}{\bibfnamefont{R.}~\bibnamefont{Prozorov}}, \bibnamefont{and}
  \bibinfo{author}{\bibfnamefont{V.~G.} \bibnamefont{Kogan}},
  \bibinfo{journal}{Phys.\ Rev.\ B} \textbf{\bibinfo{volume}{81}},
  \bibinfo{pages}{180501R} (\bibinfo{year}{2010}).

\bibitem[{\citenamefont{Blatter et~al.}(1994)\citenamefont{Blatter, Feigel'man,
  Geshkenbein, Larkin, and Vinokur}}]{Blatter94}
\bibinfo{author}{\bibfnamefont{G.}~\bibnamefont{Blatter}},
  \bibinfo{author}{\bibfnamefont{M.~V.} \bibnamefont{Feigel'man}},
  \bibinfo{author}{\bibfnamefont{V.~B.} \bibnamefont{Geshkenbein}},
  \bibinfo{author}{\bibfnamefont{A.~I.} \bibnamefont{Larkin}},
  \bibnamefont{and} \bibinfo{author}{\bibfnamefont{V.~M.}
  \bibnamefont{Vinokur}}, \bibinfo{journal}{Rev.\ Mod.\ Phys.}
  \textbf{\bibinfo{volume}{66}}, \bibinfo{pages}{1125} (\bibinfo{year}{1994}).

\bibitem[{\citenamefont{Schnack et~al.}(1993)\citenamefont{Schnack, Griessen,
  Lensink, and Hai-Hu}}]{Schnack93}
\bibinfo{author}{\bibfnamefont{H.~G.} \bibnamefont{Schnack}},
  \bibinfo{author}{\bibfnamefont{R.}~\bibnamefont{Griessen}},
  \bibinfo{author}{\bibfnamefont{J.~G.} \bibnamefont{Lensink}},
  \bibnamefont{and} \bibinfo{author}{\bibfnamefont{W.}~\bibnamefont{Hai-Hu}},
  \bibinfo{journal}{Phys.\ Rev.\ B} \textbf{\bibinfo{volume}{48}},
  \bibinfo{pages}{13178} (\bibinfo{year}{1993}).

\bibitem[{\citenamefont{Griessen et~al.}(1994)\citenamefont{Griessen, Hai-hu,
  van Dalen, Dam, Rector, Schnack, Libbrecht, Osquiguil, and
  Bruynseraede}}]{Griessen94}
\bibinfo{author}{\bibfnamefont{R.}~\bibnamefont{Griessen}},
  \bibinfo{author}{\bibfnamefont{W.}~\bibnamefont{Hai-hu}},
  \bibinfo{author}{\bibfnamefont{A.~J.~J.} \bibnamefont{van Dalen}},
  \bibinfo{author}{\bibfnamefont{B.}~\bibnamefont{Dam}},
  \bibinfo{author}{\bibfnamefont{J.}~\bibnamefont{Rector}},
  \bibinfo{author}{\bibfnamefont{H.~G.} \bibnamefont{Schnack}},
  \bibinfo{author}{\bibfnamefont{S.}~\bibnamefont{Libbrecht}},
  \bibinfo{author}{\bibfnamefont{E.}~\bibnamefont{Osquiguil}},
  \bibnamefont{and}
  \bibinfo{author}{\bibfnamefont{Y.}~\bibnamefont{Bruynseraede}},
  \bibinfo{journal}{Phys.\ Rev.\ Lett.} \textbf{\bibinfo{volume}{72}},
  \bibinfo{pages}{1910} (\bibinfo{year}{1994}).

\bibitem[{\citenamefont{Ijaduola et~al.}(2006)\citenamefont{Ijaduola, Thompson,
  Feenstra, Christen, Gapud, and Song}}]{Ijaduola06}
\bibinfo{author}{\bibfnamefont{A.~O.} \bibnamefont{Ijaduola}},
  \bibinfo{author}{\bibfnamefont{J.~R.} \bibnamefont{Thompson}},
  \bibinfo{author}{\bibfnamefont{R.}~\bibnamefont{Feenstra}},
  \bibinfo{author}{\bibfnamefont{D.~K.} \bibnamefont{Christen}},
  \bibinfo{author}{\bibfnamefont{A.~A.} \bibnamefont{Gapud}}, \bibnamefont{and}
  \bibinfo{author}{\bibfnamefont{X.}~\bibnamefont{Song}},
  \bibinfo{journal}{Phys.\ Rev.\ B} \textbf{\bibinfo{volume}{73}},
  \bibinfo{pages}{134502} (\bibinfo{year}{2006}).

\bibitem[{\citenamefont{Pratt et~al.}(2009)\citenamefont{Pratt, Tian, Kreyssig,
  Zarestky, Nandi, Ni, Bud'ko, Canfield, Goldman, and McQueeney}}]{Pratt09}
\bibinfo{author}{\bibfnamefont{D.~K.} \bibnamefont{Pratt}},
  \bibinfo{author}{\bibfnamefont{W.}~\bibnamefont{Tian}},
  \bibinfo{author}{\bibfnamefont{A.}~\bibnamefont{Kreyssig}},
  \bibinfo{author}{\bibfnamefont{J.~L.} \bibnamefont{Zarestky}},
  \bibinfo{author}{\bibfnamefont{S.}~\bibnamefont{Nandi}},
  \bibinfo{author}{\bibfnamefont{N.}~\bibnamefont{Ni}},
  \bibinfo{author}{\bibfnamefont{S.~L.} \bibnamefont{Bud'ko}},
  \bibinfo{author}{\bibfnamefont{P.~C.} \bibnamefont{Canfield}},
  \bibinfo{author}{\bibfnamefont{A.~I.} \bibnamefont{Goldman}},
  \bibnamefont{and} \bibinfo{author}{\bibfnamefont{R.~J.}
  \bibnamefont{McQueeney}}, \bibinfo{journal}{Phys.\ Rev.\ Lett.}
  \textbf{\bibinfo{volume}{103}}, \bibinfo{pages}{087001}
  (\bibinfo{year}{2009}).

\bibitem[{\citenamefont{Park et~al.}(2009)\citenamefont{Park, Inosov,
  Niedermayer, Sun, Haug, Christensen, Dinnebier, Boris, Drew, Schulz
  et~al.}}]{Park09}
\bibinfo{author}{\bibfnamefont{J.~T.} \bibnamefont{Park}},
  \bibinfo{author}{\bibfnamefont{D.~S.} \bibnamefont{Inosov}},
  \bibinfo{author}{\bibfnamefont{C.}~\bibnamefont{Niedermayer}},
  \bibinfo{author}{\bibfnamefont{G.~L.} \bibnamefont{Sun}},
  \bibinfo{author}{\bibfnamefont{D.}~\bibnamefont{Haug}},
  \bibinfo{author}{\bibfnamefont{N.~B.} \bibnamefont{Christensen}},
  \bibinfo{author}{\bibfnamefont{R.}~\bibnamefont{Dinnebier}},
  \bibinfo{author}{\bibfnamefont{A.~V.} \bibnamefont{Boris}},
  \bibinfo{author}{\bibfnamefont{A.~J.} \bibnamefont{Drew}},
  \bibinfo{author}{\bibfnamefont{L.}~\bibnamefont{Schulz}},
  \bibnamefont{et~al.}, \bibinfo{journal}{Phys.\ Rev.\ Lett}
  \textbf{\bibinfo{volume}{102}}, \bibinfo{pages}{117006}
  (\bibinfo{year}{2009}).

\bibitem[{com()}]{comment}
\bibinfo{note}{Instead of theoretically proposed values $p=2$\ and $n=5/2$, in
  Ref.$^2$ the critical exponents $p=1/4$, $n=1$\ and $p=1$, $n=3/2$\ have been
  used in the analysis of the critical current density $j_c$. We note, however,
  that the plot of Eq.(4) with theoretical values $p=2$\ and $n=5/2$\ is rather
  similar to the plot of this function with the parameters used in Ref.$^2$.}

\bibitem[{\citenamefont{van~der Beek et~al.}(2010)\citenamefont{van~der Beek,
  Rizza, Konczykowski, Fertey, Monnet, Klein, Okazaki, Ishikado, Kito, Iyo
  et~al.}}]{Beek10}
\bibinfo{author}{\bibfnamefont{C.~J.} \bibnamefont{van~der Beek}},
  \bibinfo{author}{\bibfnamefont{G.}~\bibnamefont{Rizza}},
  \bibinfo{author}{\bibfnamefont{M.}~\bibnamefont{Konczykowski}},
  \bibinfo{author}{\bibfnamefont{P.}~\bibnamefont{Fertey}},
  \bibinfo{author}{\bibfnamefont{I.}~\bibnamefont{Monnet}},
  \bibinfo{author}{\bibfnamefont{T.}~\bibnamefont{Klein}},
  \bibinfo{author}{\bibfnamefont{R.}~\bibnamefont{Okazaki}},
  \bibinfo{author}{\bibfnamefont{M.}~\bibnamefont{Ishikado}},
  \bibinfo{author}{\bibfnamefont{H.}~\bibnamefont{Kito}},
  \bibinfo{author}{\bibfnamefont{A.}~\bibnamefont{Iyo}}, \bibnamefont{et~al.},
  \bibinfo{journal}{Phys.\ Rev.\ B} \textbf{\bibinfo{volume}{81}},
  \bibinfo{pages}{174517} (\bibinfo{year}{2010}).

\bibitem[{\citenamefont{Pippard}(1969)}]{Pippard69}
\bibinfo{author}{\bibfnamefont{A.~B.} \bibnamefont{Pippard}},
  \bibinfo{journal}{Philos.\ Mag.} \textbf{\bibinfo{volume}{19}},
  \bibinfo{pages}{217} (\bibinfo{year}{1969}).

\bibitem[{\citenamefont{Larkin and Ovchinnikov}(1979)}]{Larkin79}
\bibinfo{author}{\bibfnamefont{A.~I.} \bibnamefont{Larkin}} \bibnamefont{and}
  \bibinfo{author}{\bibfnamefont{Y.~N.} \bibnamefont{Ovchinnikov}},
  \bibinfo{journal}{J.\ Low\ Temp.\ Phys.} \textbf{\bibinfo{volume}{34}},
  \bibinfo{pages}{409} (\bibinfo{year}{1979}).

\bibitem[{\citenamefont{Gordon et~al.}(2009{\natexlab{b}})\citenamefont{Gordon,
  Ni, Martin, Tanatar, Vannette, Kim, Samolyuk, Schmalian, Nandi, Kreyssig
  et~al.}}]{Gordon09b}
\bibinfo{author}{\bibfnamefont{R.~T.} \bibnamefont{Gordon}},
  \bibinfo{author}{\bibfnamefont{N.}~\bibnamefont{Ni}},
  \bibinfo{author}{\bibfnamefont{C.}~\bibnamefont{Martin}},
  \bibinfo{author}{\bibfnamefont{M.~A.} \bibnamefont{Tanatar}},
  \bibinfo{author}{\bibfnamefont{M.~D.} \bibnamefont{Vannette}},
  \bibinfo{author}{\bibfnamefont{H.}~\bibnamefont{Kim}},
  \bibinfo{author}{\bibfnamefont{G.~D.} \bibnamefont{Samolyuk}},
  \bibinfo{author}{\bibfnamefont{J.}~\bibnamefont{Schmalian}},
  \bibinfo{author}{\bibfnamefont{S.}~\bibnamefont{Nandi}},
  \bibinfo{author}{\bibfnamefont{A.}~\bibnamefont{Kreyssig}},
  \bibnamefont{et~al.}, \bibinfo{journal}{Phys.\ Rev.\ Lett.}
  \textbf{\bibinfo{volume}{102}}, \bibinfo{pages}{127004}
  (\bibinfo{year}{2009}{\natexlab{b}}).


\end{thebibliography}
\end{document}